\documentclass[11pt,preprint]{aastex}
\usepackage{rotating,epsfig,natbib}

\slugcomment{to appear in the Astrophysical Journal}

\shorttitle{The Broadband Emission Spectrum of HD~189733b}
\shortauthors{Charbonneau et al.}

\def\simgr{\,\hbox{\hbox{$ > $}\kern -0.8em \lower 1.0ex\hbox{$\sim$}}\,}
\def\simle{\,\hbox{\hbox{$ < $}\kern -0.8em \lower 1.0ex\hbox{$\sim$}}\,}

\begin{document}

\title{The Broadband Infrared Emission Spectrum of the Exoplanet HD~189733b}

\author{David Charbonneau\altaffilmark{1,2,3}, Heather A. Knutson\altaffilmark{1}, 
Travis Barman\altaffilmark{4}, Lori E. Allen\altaffilmark{1}, Michel Mayor\altaffilmark{5}, 
S. Thomas Megeath\altaffilmark{6}, Didier Queloz\altaffilmark{5}, 
and St\'{e}phane Udry\altaffilmark{5}}
\altaffiltext{1}{Harvard-Smithsonian Center for Astrophysics, 60 Garden St., Cambridge, MA 02138}
\altaffiltext{2}{dcharbonneau@cfa.harvard.edu}
\altaffiltext{3}{Alfred P. Sloan Research Fellow}
\altaffiltext{4}{Lowell Observatory, 1400 W. Mars Hill Rd., Flagstaff, AZ 86001}
\altaffiltext{5}{Observatoire de Gen\`eve, Universit\'e de Gen\`eve, 51 ch. des Maillettes, 1290 Sauverny, Switzerland}
\altaffiltext{6}{Department of Physics and Astronomy, University of Toledo, 2801 West Bancroft St., Toledo, OH, 43606}

\begin{abstract}
We present {\it Spitzer Space Telescope} time series photometry of the exoplanet system HD~189733 spanning two times of
secondary eclipse, when the planet passes out of view behind the parent star.  We estimate the relative eclipse depth in 5 distinct bands and 
find the planet-to-star flux ratio to be $0.256 \pm 0.014$\% (3.6 \micron), $0.214 \pm 0.020$\% (4.5 \micron), 
$0.310 \pm 0.034$\% (5.8 \micron), $0.391 \pm 0.022$\% (8.0 \micron), and $0.598 \pm 0.038$\% (24 \micron).  For consistency, we
re-analyze a previously published time series to deduce a contrast ratio in an additional band, $0.519 \pm 0.020$\% (16 \micron).
Our data are strongly inconsistent with a Planck spectrum, and we clearly detect emission near 4 \micron\ as predicted
by published theoretical models in which this feature arises from a corresponding opacity window.
Unlike recent results for the exoplanet HD~209458b, we find that the emergent spectrum from HD~189733b is 
best matched by models that do not include an atmospheric temperature inversion.  Taken together, these two studies provide initial 
observational support for
the idea that hot Jupiter atmospheres diverge into two classes, in which a thermal inversion layer is present for the more
strongly irradiated objects.
\end{abstract}


\section{Introduction}
Of the many intriguing avenues for investigation afforded by transiting exoplanets 
\citep[for a review, see][]{charbonneau2007},
perhaps the most exciting is the opportunity to study directly the planetary atmosphere
without the need to spatially resolve the light from the planet from that of its host star.  At visible
wavelengths, where the emitted light from the planet is negligible, the technique of transmission spectroscopy 
has been employed from space-based platforms to study atoms, molecules, and clouds in the atmospheres of HD~209458b 
\citep{charbonneau2002, vidal2003, barman2007} and HD~189733b \citep{tinetti2007, pont2008, swain08a}.
Ground-based studies had yielded only upper limits, albeit valuable ones \citep{moutou2001, moutou2003, 
winn2004, deming2005a, narita2005, arribas2006, bozorgnia2006} until the recent breakthrough detection by
\cite{redfield2008} of atomic sodium in the atmosphere of HD~189733b.  
Searches for reflected starlight have delivered increasingly stringent constraints on the 
wavelength-dependent geometric albedo and phase function \citep{charbonneau1999,collier2002,leigh2003a,leigh2003b,
rowe2006,rowe2008}.  At infrared wavelengths, the planet-to-star contrast ratio improves dramatically, permitting a direct study
of the emitted planetary radiation through the modulation of the total light from the star and planet
spanning times of secondary eclipse, when the planet is occulted by the star.  Ground-based searches
for this effect have been frustrated to date by telluric variability 
\citep{richardson2003a, richardson2003b, snellen2005, deming2007a, knutson07b}, although \cite{snellen2007} have recently claimed
a tentative detection for OGLE-TR-113b.   Exemplifying the power of a small-aperture telescope
in an ideal location, the {\it Spitzer Space Telescope} \citep{werner2004} has revolutionized
this field by yielding a flurry of detections
of the emitted radiation from 6 exoplanets \citep{charbonneau2005, deming2005b, 
deming2006a, deming2007b, demory2007, harrington2006, harrington2007, knutson07a, knutson08}, 
including recent determinations of the emitted spectra \citep{grillmair2007, richardson2007, swain08b}
of HD~189733b and HD~209458b.
Numerous ongoing programs promise additional exciting results to come.  Although {\it Spitzer} will
exhaust its cryogen in the spring of 2009, two photometric channels will continue to operate at full strength
and will provide a singular window into the atmospheres of these distant worlds \citep{deming2007c},
if the warm phase of the {\it Spitzer} Mission is approved.

Each of the techniques described above is facilitated for large planets that orbit nearby, bright stars
at small orbital separations.  The discovery of the exoplanet HD~189733b \citep{bouchy2005} provided a 
golden opportunity for practitioners of such techniques for two reasons.  First, it is at a distance of only 19.3~pc, 
and hence at infrared wavelengths it is the brightest star ($K=5.5$) known to host a transiting exoplanet.
Second, the planet-to-star surface area ratio is the second largest of the 28 transiting exoplanets described
in the literature, surpassed by only TrES-3 \citep{odonovan2007}.
It is primarily for these reasons of accessibility that we selected HD~189733b for intensive study with {\it Spitzer}.
For the purposes of this paper, we adopt the values for orbital inclination $i$, the planet
radius $R_p$, the stellar radius $R_{\star}$, and the time of transit center $T_c$ as determined by \cite{knutson07a},
and the orbital period $P$ as determined by \cite{winn07a}.
We note that the target star has an M-dwarf companion at a projected physical
separation of 216 AU \citep{bakos2006}.  This companion appears in the field of
view of the data we present below, and we comment upon its impact upon our
photometric results when appropriate.

The recent {\it Spitzer} detections of the flux emitted by several exoplanets have spurred 
numerous theoretical efforts to model their atmospheres and predict their emitted spectra \citep[see for example][]{barman2005,
burrows2005, burrows2006, fortney2005, fortney2006, seager2005}.  The constraints on the theoretical predictions 
become much more stringent with detections at numerous wavelengths, from which a 
broadband spectrum may be inferred.  In this paper, we describe our estimate of planet-to-star 
contrast ratio in bands centered at $3.6, 4.5, 5.8, 8.0, {\rm and}\, 24\, {\mu}m$.  We then
combine these results with our re-analysis of the earlier measurement at $16\, {\mu}m$ \citep{deming2006a}
to consider what constraints may be placed on the temperature and chemistry of the planetary 
atmosphere, and the efficiency with which it redistributes the incident stellar radiation
from the day side to the night side.  These observations are particularly timely, given the recent 
study by \citet{knutson08}, which included data presented in Deming et al.\ (2005b), of the broad band 
spectrum of HD~209458b in 5 of the above bandpasses.
In the assembled spectrum, features that had been predicted in absorption were observed in emission, a clear sign of an atmospheric
temperature inversion \citep{burrows2007a}.  \cite{fortney2008} and \cite{burrows2008} have recently proposed a division between the 
predicted atmospheric temperature-pressure profiles for hot Jupiters dependent upon the level of stellar 
insolation, in which only the most strongly irradiated planets show a temperature inversion.  
According to the criterion proposed by \cite{fortney2008}, HD~209458b is a member
of the pM class and hence is expected to show a temperature inversion, whereas HD~189733b is a member of the pL class and should not.  
Intriguingly, HD~209458b and HD~189733b also differ in that the radius of the former \citep[$1.32\, R_{\rm Jup}$;][]{knutson07c} is much larger than 
that of the latter \citep[$1.14\, R_{\rm Jup}$;][]{knutson07a} and is at odds with the predictions of models 
that do not include an additional energy source interior to the planet.  Whether the large radius and the existence of a 
temperature inversion share a common physical origin is unclear, but surely worthy of further investigation.
Our study of the broad band spectrum of HD~189733b provides an ideal opportunity to conduct a comparative study with HD~209458b.

\section{Observations and Analysis}\label{obs}

\subsection{IRAC Observations}

We observed HD 189733b over a period of 4.9  hours on UT 2005 November 24, spanning a single secondary eclipse, using the Infrared Array Camera \citep[IRAC;][]{fazio2004} on the \emph{Spitzer Space Telescope}.  We observed in subarray mode with an exposure time of 0.1~s and cycled between the four IRAC channels in order to obtain estimates of the depth of the eclipse at 3.6, 4.5, 5.8, and 8.0 \micron~simultaneously.  We obtained a total of 21,760 $32 \times 32$ pixel images in each channel\footnote{We use images processed using version S13.0 of the standard Spitzer pipeline, to avoid the additional noise introduced by a dark drift correction that was applied to all subarray images beginning with version S14.0 of the pipeline, released in May 2006.  This dark drift correction is poorly constrained for subarray images dominated by a single bright star, and as a result introduces noise at a level higher than the effect it is meant to correct.  See IRAC pipeline history available online at http://ssc.spitzer.caltech.edu/irac/ for more information.  This is important for only the 3.6~\micron~channel, where the eclipse depth is partially degenerate with the decorrelation as a function of x and y position on the array.  If we use the new version of the pipeline with the poorly constrained dark drift correction, we find the measured x and y positions of the star on the array are shifted by up to 0.001 pixels and the depth of the eclipse in this channel increases to 0.0041.}.  Images are taken in sets of 64, and four sets of images ($4 \times 64$ images total) are obtained in each channel before re-pointing the telescope to position the star on the subarray for the next channel.  The position of the star on the subarray is still varying through the first set of 64 images after each re-pointing, and we chose to discard this initial set of images, leaving a total of 16,320 usable images in each of the four channels.  The total size of this pointing drift is 0.3 pixels in the first set of 64 images, and 0.1 pixels or less in the following three sets of images.  For each image, we calculate the JD value for the time at mid-exposure, and apply a correction to convert these JD values to the appropriate HJD taking into account the position of {\it Spitzer} in the Solar system at each point during the observations.  

Because the two shortest wavelength IRAC channels (3.6 and 4.5~\micron) use InSb detectors and the two longer wavelength channels (5.8 and 8.0~\micron) use Si:As detectors, there are fundamental differences between the properties of the data taken with these two types of detectors.  We describe our analysis for each type of detector separately below.  The specific methods we employ our similar to those described in \cite{knutson08}.

\subsubsection{3.6 and 4.5~\micron~Observations (InSb Detector)}\label{short_norm}

Because HD 189733 is a bright star and the background at these shorter wavelengths is minimal, we calculate the flux from the star in each image using aperture photometry with a circular aperture with a radius of five pixels.  We determine the position of the star in each image as the position-weighted sum of the flux in a $7\times7$ pixel box centered on the approximate position of the star.  We estimate the background in each image by selecting a subset of pixels from the corners of the image where the point spread function of the star is faintest and avoiding the region around the M dwarf companion, making a histogram of the flux values in these pixels, and fitting a Gaussian function to the center of this distribution.

Fluxes measured at these two wavelengths show a strong correlation with the changing position of the star on the array, at a level comparable to the depth of the secondary eclipse.  This effect is due to a well-documented intra-pixel sensitivity \citep{reach05,charbonneau2005,mor06,knutson08}, and can be removed by fitting the data with a quadratic function of the sub-pixel position of the center of light of the stellar image, according to:
\begin{equation}
f^1 = f\, [ c_1 + c_2\, ( x - 13.5 ) + c_3\, ( x - 13.5 )^2 + c_4\, ( y - 14.5 ) + c_5\, ( y - 14.5 )^2 ]
\end{equation}
where $f$ is the flux from the star prior to this distortion, $f^1$ is the measured flux, $x$ and $y$ denote the location in pixels of the center of the star on the subarray, and $c_1-c_5$ are the five free parameters in the fit. 

\begin{figure}[h]
\epsscale{0.75}
\plotone{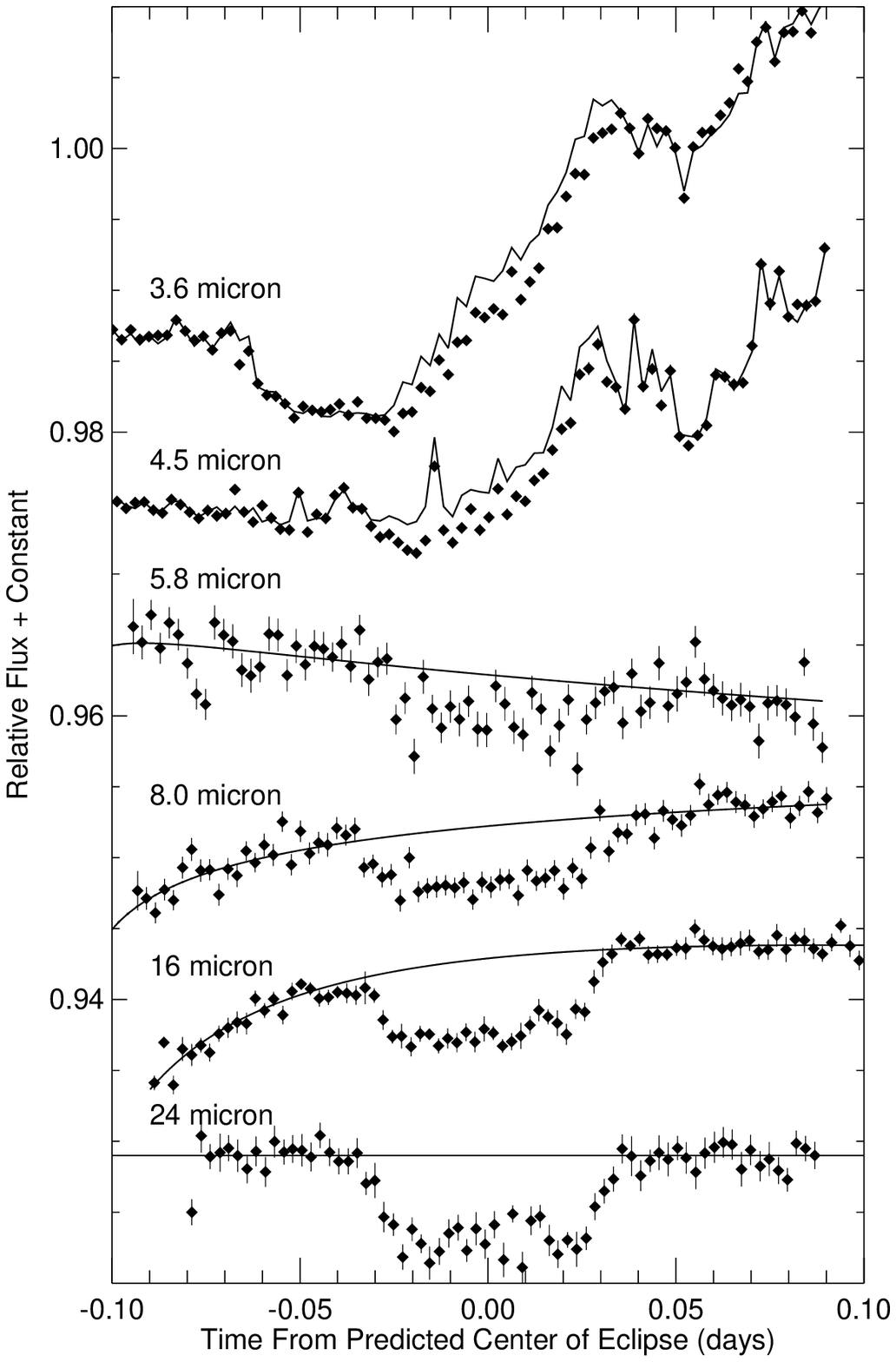}
\caption{Raw time series observations of HD 189733b spanning predicted times of secondary eclipse.  The central wavelength of observation for each data set is (from top to bottom) 3.6, 4.5, 5.8, 8.0, 16, and 24 \micron. Each time series is binned in 3.5 minute intervals, normalized, and plotted with a distinct constant offset for clarity.  The overplotted curves show the best-fit corrections for detector effects (see \S\ref{obs}).
\label{eclipse_plot_no_norm}}
\end{figure}

\begin{figure}[h]
\epsscale{0.75}
\plotone{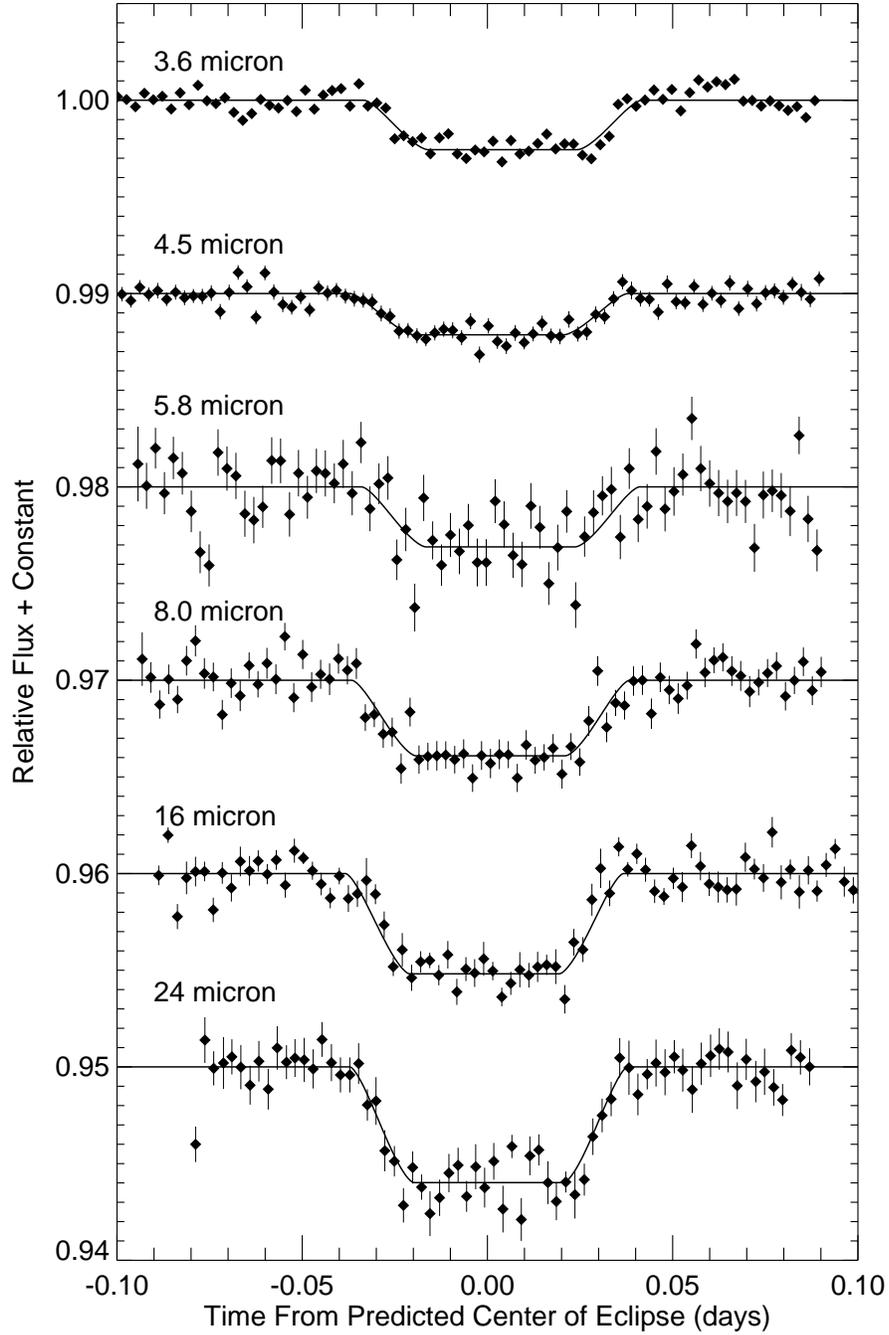}
\caption{Time series observations of HD 189733b after correcting for detector effects (see \S\ref{obs}).  The central wavelength of observation for each data set is (from top to bottom) 3.6, 4.5, 5.8, 8.0, 16, and 24 \micron.  Each time series is binned in 3.5 minute intervals, normalized, and plotted with a distinct constant offset for clarity.  The best-fit eclipse curves are overplotted.
\label{eclipse_plot_with_norm}}
\end{figure}

We fit the correction for the intra-pixel sensitivity of the array and the eclipse curve simultaneously to the data using a Markov Chain Monte Carlo method \citep[see for example][]{ford05,winn07b} with $10^6$ steps.  We set the uncertainty on individual points equal to the standard deviation of the out-of-eclipse data after correction for the intra-pixel variations, and remove outliers of 5$\sigma$~or more as calculated using the residuals from the best-fit light curve.  We allow both the depth and timing of the secondary eclipse to vary independently for the eclipses at each of the two observed wavelengths, and fix the values of $i$, $R_p$, $R_{\star}$, and $T_c$ to those given in \citet{knutson07a}, the value for $P$ to that found by \citet{winn07a}.  We calculate our eclipse curve using the equations from \citet{mand02} for the case with no limb-darkening.  After running the chain, we search for the point in the chain where the $\chi^2$ value first falls below the median of all the $\chi^2$ values in the chain (i.e.\ where the code had first found an excellent fit), and discard all the steps up to that point.  We take the median of the remaining distribution as our best-fit parameter, with errors calculated as the symmetric range about the median containing 68\% of the points in the distribution.  The distribution of values was very close to symmetric in all cases, and there did not appear to be any strong correlations between variables.  Figure \ref{eclipse_plot_no_norm} shows the binned data with the best fit to the detector effects overplotted, and Figure \ref{eclipse_plot_with_norm} shows the binned data once these trends are removed, with best-fit eclipse curves overplotted.  The best-fit eclipse depths and times are given in Table \ref{eclipse_depths}.

\subsubsection{5.8 and 8.0~\micron~Observations (Si:As Detector)}\label{long_norm}

At longer wavelengths the flux from the star is smaller and the zodiacal background is larger; as a result we chose to use a smaller circular aperture with a radius of 3.5 pixels in order to minimize the noise contribution from this increased background.  As before, we calculate the position of the star individually in each image as the position-weighted sum of the fluxes in a $7\times7$ pixel box, and estimate the background  using a Gaussian fit to a histogram of the pixels in the corners of the array.  Fluxes in the first 10 images and the 58th image in each set of 64 are consistently below the median value for the set by as much as 10\%, with the lowest values at the beginning of each set, so we chose to exclude these 11 images from each set of 64 in our analysis.  

There is no known intra-pixel sensitivity at these wavelengths \citep{charbonneau2005}, but there is another well-documented detector effect \citep{knutson07a}, which causes the effective gain (and thus the measured flux) in individual pixels to increase over time.  This effect has been referred to as the ``detector ramp'', and has also been observed \citep{deming2006a} in the IRS 16~\micron~array, which is made from the same material.  The size of this effect depends on the illumination level of the individual pixel, such that the time-dependent gain of different pixels form a well-defined family of functions.  The most strongly-illuminated pixels ($>$250 MJy Sr$^{-1}$ in the 8~\micron~channel) converge asymptotically to a constant value within the first hour of observations.
The time scale for this convergence increases with decreasing illumination, and the limiting case (for pixels that receive little illumination) is that
of a linear increase in the measured flux over time, with a slope that varies inversely with the logarithm of the illumination level.

This effect is important for two reasons.  First, it means that the observed 3\% increase in the measured background flux at 8~\micron~over the period of the observations is most likely not the result of a real change in the zodiacal background, but is instead another example of this detector ramp. Although the increased noise and smaller size of the background at 5.8~\micron~obscures this effect, there appears to be a similar upward trend.  Thus, rather than calculating the background in each image individually and subtracting that value, we subtract a constant background of 4.81 MJy Sr$^{-1}$ per pixel from all the 8~\micron~images, and 0.96 MJy Sr$^{-1}$ per pixel from all 5.8~\micron~images.  This background is calculated as the median background value during the last hour of observation, when presumably the background is closest to the value it would ultimately obtain.  We note that this choice has a negligible effect on our final eclipse depths, as the background constitutes only 0.2\% and 1.7\% of the signal in our aperture at 5.8 and 8.0~\micron, respectively.  

This effect also produces a 1\% increase in the measured flux from the star at 8.0~\micron~over the period of these observations (see Figure~1).  There does not appear to be any ramp visible at 5.8~\micron; instead this data shows a general downward trend.  Unlike the detector ramp at low illumination levels, the ramp for higher illuminations has an asymptotic shape, with a steeper rise in the first 30 minutes of observations.  We discard the first 30 minutes of data in both the 5.8 and 8.0~\micron~channels and fit the remaining binned time series from our 3.5 pixel aperture with a quadratic function of $\ln{{\Delta} t}$, where ${\Delta} t$ is the change in time from the start of the observations.  \citet{knutson07a} used this same functional form (with additional degrees of freedom) to describe the detector ramp in the 8~\micron~channel over a period of 33 hours of continuous observations of HD~189733, and it accurately captures the behavior of this ramp for a range of illumination levels.  Unlike \citet{knutson07a}, we do not attempt to correct each of the pixels in the images individually for this ramp, as this is not necessary for our analysis.  Moreover, the lower fluxes, shorter time frame (5 hours instead of 33) and reduced cadence of our data (from cycling between the four detectors) make it difficult to characterize this effect accurately at the pixel level.  Instead we assume the effect of detector ramp in the binned flux in our 3.5 pixel aperture will have a shape that is representative of the flux-weighted average of the ramp effect for individual pixels.

We fit both the quadratic function of $\ln{{\Delta} t}$  and the theoretical eclipse curve to the data simultaneously using a Markov Chain Monte Carlo method as described in \S\ref{short_norm}.  As before, the distribution of values was very close to symmetric in all cases, and there did not appear to be any strong correlations between variables.  Best-fit eclipse depths and times from these fits are given in Table \ref{eclipse_depths}, and the time series both before and after correcting for detector effects are shown in Figures  \ref{eclipse_plot_no_norm} and \ref{eclipse_plot_with_norm}, respectively.  

\subsection{IRS Photometry}

As a check and to ensure a consistent methods between different bandpasses, we re-analyze the raw time series gathered with the {\it Spitzer} IRS instrument (Houck et al.\ 2004) at 16~\micron~(prior to normalization using the time series for the M dwarf companion) from \citet{deming2006a}.  Importantly, these data were obtained on UT 2005 November 17, which falls between the times
of our observations with MIPS (\S2.3) and IRAC (\S2.1).  We fit the 16 \micron\  time series with the same quadratic function of $\ln{{\Delta} t}$ as the 5.6 \micron\ and 8.0~$\mu$m data.  As before, we fit this normalization and the eclipse curve to the data simultaneously using a Markov Chain Monte Carlo method with the error for individual points set equal to the standard deviation of the points after the end of the eclipse, calculated after fitting this section of data with a linear function of time to remove the overall slope.  This data shows the same type of detector ramp as the 8.0~\micron~data, and we exclude from our fit the first 30 minutes of data where this ramp is steepest.  Our estimate of the eclipse depth, $0.519 \pm 0.020$\%, is slightly shallower and has a smaller uncertainty than the value reported by \citet{deming2006a}, $0.551 \pm 0.030$\%.  We also performed a separate analysis of the time series assuming a simple quadratic function of time, which we found yielded the same eclipse depth and uncertainty as that stated in Deming et al.\ (2006).  We note that the analysis performed by those authors differed from ours in that they first divided the time series of HD~189733 by a fourth-order polynomial-fit in $\ln{{\Delta} t}$ to the time series of the M-dwarf companion, and subsequently fit a quadratric function of time to the out-of-eclipse portion of the corrected time series.  They also reported the results of a linear fit to the corrected time series, for which they obtained an eclipse depth of 0.521\%, very similar to our final value, stated above.  As discussed in \S\ref{long_norm}, we feel that fitting with a quadratic function of $\ln{{\Delta} t}$ represents both a more robust normalization than a quadratic function of time, as well as being a better description of the shape of the detector ramp.  The reduced $\chi^2$ is 1.07 for the fit that uses a quadratic function of time, and falls to 1.00 when we adopt a quadratic function of $\ln{{\Delta} t}$, indicating that the latter function is indeed a better fit to the data.  Importantly, the difference in eclipse depths between the two normalizations is less than $1\sigma$.

\subsection{MIPS Observations}

We observed HD~189733 for 5 hours spanning a time of secondary eclipse on UT 2005 November 11, and obtained 1548 images using the \emph{Spitzer} MIPS 24~$\mu$m array \citep{rieke2004}.  The MIPS observing sequence dithers the target position amongst 14 distinct positions on the detector array (see \S8.2.1.2.1 of the \emph{Spitzer Observer's Manual}).  Due to the variations in apparent sensitivity at each position, we elected to treat the series of images at each position as independent data sets.  After we produce a photometric time series for each position, we normalize each time series to a common scale and subsequently combine them to form a single time series, as described below.

We estimate the sky background in each image from a \mbox{40 $\times$ 40} pixel box centered on the star, trimming all pixels more than $3\sigma$ away from the median value and fitting a Gaussian function to the central region of a histogram of the remaining pixels.  We find the average background over the duration of our observations is $21.9\ {\rm MJy\, sr^{-1}}$.  We then subtract this background from each image individually, and fit the remaining flux in a \mbox{10 $\times$ 10} pixel subarray (which is large enough to encompass the first Airy ring) centered on the approximate position of the star.  We found that varying the size of the subarray we used did not affect the final time series.  We use a MIPS model point spread function (PSF) for a 5000~K point source\footnote{Available at http://ssc.spitzer.caltech.edu/mips/psf.html} and allow both the $x$ and $y$ position of the PSF as well as a constant scaling factor to vary in our fits.  We correct for the effects of transient hot pixels in our subarray by making a histogram of flux values at a given pixel position over the set of images (of which there were $108-126$) at a given nod position and flag $4\sigma$ outliers.  We then assign a statistical weight of zero to these values in the fitting process described above; $95\%$ of our images have one or fewer of these hot pixels.  We discard images with 5 or more hot pixels in the subarray; there are only 10 of these images in the entire time series.  We note that the M-dwarf companion to HD 189733, is included within our 10 pixel subarray; we give zero statistical weight to the values within a 3 $\times$ 3 pixel box centered on the position of the companion in our fits.

We note that the measured background periodically drops by $1.5\%$, corresponding to the first image in each bundle of images spanning the sequence of nod positions.  \cite{deming2005b} noted the same feature in their data, and chose to divide the measured flux by the background in order to remove the effect of this drop.  This procedure would introduce significant noise in our data, because (unlike the case of HD~209458) our target star has a flux twice the value of the background. Thus we elected to simply discard the 136 images with these low sky backgrounds.  

For times later than 1.5 hours after the start of observations, the final time series at each of the 14 nod positions is well-described by a combination of the theoretical eclipse curve and a linear function of time.  At earlier times, there appears to be a slight upward trend, possibly corresponding to the detector ramp described in \S\ref{long_norm}.  We fit the entire time series, but find that clipping this initial segment of the data in our fits changes the final measured eclipse depth by only $0.5\sigma$.  We fit a single eclipse curve and 14 linear functions of time (one for each of the 14 nod positions) to the data simultaneously using a Markov Chain Monte Carlo method as described in \S\ref{short_norm}, trimming $3\sigma$ outliers in our fit.   A binned version of the combined time series appears in Figures \ref{eclipse_plot_no_norm} and \ref{eclipse_plot_with_norm}, and we state our estimates of the eclipse depth and its time of center in Table~1.

\section{Discussion and Conclusions}

\begin{figure}[ht]
\epsscale{1.}
\plotone{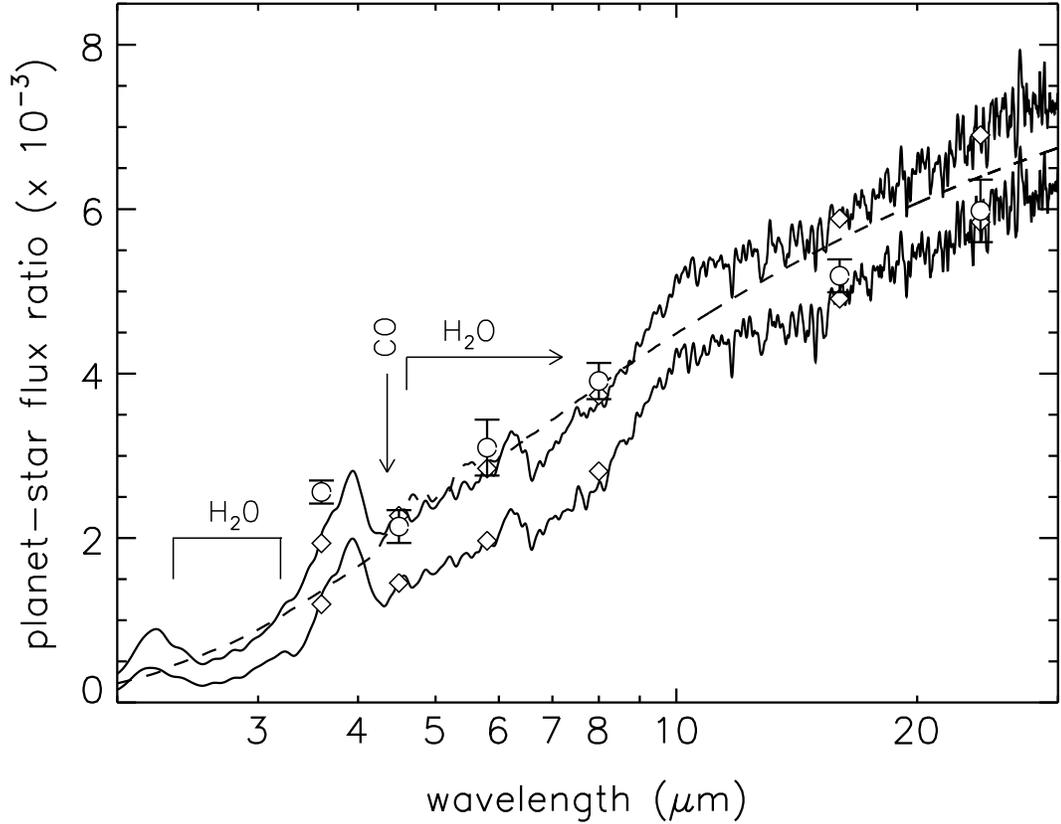}
\caption{
Comparison between IRAC (3.6, 4.5, 5.8, \& 8.0 \micron), IRS (16 \micron), and MIPS (24 \micron) eclipse depths (circles) and model
planet-star flux ratios (solid lines).  The top black curve is a model in which the 
emission of the absorbed stellar flux is constrained to the day side only.  The lower black curve is
a model with uniform energy redistribution over the entire planet.
Integration of these synthetic spectra over the $Spitzer$ band-passes are
indicated with diamonds.  The flux ratio expected under the assumption that the planet
radiates a Planck spectrum with a temperature of 1292~K planet is shown as the dashed line.
\label{modelfig}}
\end{figure}

With the 6 distinct band-integrated flux ratio measurements in hand, we proceeded
to compare these values with model predictions.  Following Barman et al.\ (2005), two model 
atmospheres for HD~189733b were constructed, one assuming that the absorbed stellar flux is confined to 
the day side while the other model uniformly distributes the stellar flux over the
entire planet.  Figure \ref{modelfig} compares these models and the best-fit Planck curve
(corresponding to a temperature of 1292~K) to the measured day side
secondary eclipse depths.  We refer the reader
to Barman (2008) for a more detailed examination of atmospheric models of
HD~189733b in light of our data, but we discuss several of the salient points below.

Our data show clear deviations from the Planck curve.
The greatest discrepancy occurs at 3.6~\micron, for which the planet-to-star contrast ratio is 
more than twice as great as would be expected from pure blackbody emission.  This emission feature 
is a prediction that is common to nearly all published theoretical models (e.g.\ Barman et al.~2005; Burrows et al.\ 2006; 
Fortney et al.\ 2006; Seager et al.\ 2005).  It arises from a hole in the atmospheric opacities near 4~\micron;
significant absorption from both water and carbon monoxide in adjacent wavelengths regions causes flux to be squeezed out through
this wavelength span of relatively low opacity.  Furthermore, the data exhibit a clear trough from $4.5-8.0~\mu$m, which
is consistent with water absorption.  This finding is at odds with the results
from Grillmair et al.\ (2007), who found that the spectrum of HD~189733b was remarkably flat 
across the 7.5$-$14.7~\micron\ bandpass and, in particular, that it did not show the
downturn at blue wavelengths expected from water absorption.  However, the 8.0 and 16 \micron\
photometric data in Figure~3 are themselves only modestly inconsistent with the 
relatively noisy spectrum from Grillmair et al.\ (2007).  It may be that while water is indeed present and seen
in absorption at shorter wavelengths, additional spectral features are present in the $8-10$~\micron\ region
that mask the expected absorption due to water.  Fortney \& Marley (2007) have noted that the Grillmair et al.\ (2007) 
result is itself difficult to reconcile with previously published 8.0 \micron\ photometry (Knutson et al.\ 2007a),
since it would require the presence of physically implausible opacity sources.

The presence of water in absorption is in stark contrast to recent findings 
that water is in emission for HD 209458b (Knutson et al.\  2008; Burrows et al.\ 2007a)
suggesting fundamental atmospheric differences between these two planets.
Several authors (e.g. Fortney et al.\ 2008; Burrows et al.\ 2008; see also Hubeny et al.\ 2003) have 
discussed how the atmospheres of hot Jupiters might bifurcate into two
groups, with the presence or absence of a temperature inversion (as deduced from
the presence or absence of emission features in the spectrum) arising under different levels of external 
irradiation.  As noted by these authors, a candidate for the physical origin of this inversion is the presence of
gaseous TiO and VO in the upper atmospheres of the more strongly irradiated planets.  Interestingly, HD~209458b has an equilibrium temperature
of $1449 \pm 12$~K and hence is significantly more strongly illuminated than HD~189733b, which has an equilibrium temperature
of only $1201 \pm 12$~K (Torres et al.\ 2008).  While it is surely too early to state that the
presence of a thermal inversion in the atmosphere of HD~209458b and the apparent lack of such an
inversion in HD~189733b indeed result from their differing levels of irradiation, the connection is
intriguing and worthy of further study.  Moreover, not only do the overall bolometric irradiances of the two
planets different, but the G0V spectral type of HD~209458 ensures a substantially bluer stellar spectral
energy distribution than that of the K2V spectral type of HD~189733.  We speculate that the enhanced short-wavelength irradiation 
for HD~209458b might generate dramatically different photochemical products and hazes than are present on HD~189733b
(see Marley et al.\ 2007 for a discussion of photochemical hazes on hot Jupiters). Since the
presence of hazes and their approximate size distribution may be inferred from transmission spectroscopy (e.g.\
Pont et al.\ 2008), we urge future modeling efforts to consider the constraints offered by those studies in
conjunction with the emission spectroscopy results we present here.  We note that within the Solar system, planetary
atmospheres exhibit a wide variety of photochemical products and distinct physical mechanisms for temperature inversions,
and we should expect no lesser diversity amongst the atmospheres of planets orbiting other stars.  

The planets also differ substantially in terms of their physical structure:
HD~209458 is larger but less massive than HD~189733, such that the densities of the two planets,
$0.34 \pm 0.02\ {\rm g\, cm^{-3}}$ for HD~209458b and $0.96 \pm 0.08\ {\rm g\, cm^{-3}}$ for HD~189733 (Torres et al.\ 2008) differ by
nearly a factor of 3.  The larger-than-predicted radii of some, but not all gas-giant exoplanets is
one of the greatest conundrums in the field today.  Many authors have sought to explain the
radii of the puffy planets by invoking an additional source in the energy budget of those
planets (e.g. Bodenheimer et al.\ 2003; Showman \& Guillot 2002; but see Burrows et al.\ 2007b for 
a dissenting opinion).  As the atmosphere is the energy gateway for the planet, it is natural to ponder whether
there exists a physical connection between the observed properties of the atmospheres of these planets
and the significant differences in their physical structure.  The best avenue for progress on the observational front
is to gather data for a planet that represents an extreme member of one of these two classes.  We note that
TrES-4 (Mandushev et al.\ 2007) has the lowest density ($0.21 \pm 0.03\ {\rm g\, cm^{-3}}$) of the known transiting
exoplanets, and, with an equilibrium temperature of 1785~K it is also one of the most strongly irradiated.  Hence 
it may be the ideal target with which to confirm or refute this proposed connection.

The $3.6-8.0~\mu$m IRAC observations clearly favor a model that assumes no redistribution of extrinsic energy to the night side
(upper curve in Figure~3), which would predict very cold night side temperatures.  This inference is in contradiction
to the recent 8 $\mu$m light curve measurement from Knutson et al.\ (2007a).  Those authors determined that the minimum hemisphere-integrated
brightness temperature was $973 \pm 33$~K and the maximum hemisphere-integrated brightness temperature was not much
larger, $1212 \pm 11$~K.  This relatively modest day-night contrast indicated that the planet was efficiently
transporting energy from the day side to the night side.  The essence of the problem is one of energy budgeting:
Our IRAC data indicate a very hot dayside, and the Knutson et al.\ (2007a) values indicate a relatively flat phase curve.
If one assumes that the planet exhibits a similar lack of a day-night contrast at all wavelengths, the overall
emission will exceed the energy it absorbs even under the most favorable case (an albedo of 0).
Although our 16 and 24 \micron\ values are in agreement with a model with a small day-night variation
(the lower curve in Figure~3), the emission at these long wavelengths represents only a small fraction of the bolometric
flux of the planet and thus cannot, in itself, alleviate this concern.  
A partial reconciliation may be that the degree of energy redistribution is depth-dependent, and hence the
emission phase curve of the planet is strongly wavelength dependent, such that the planet radiates
relatively little energy from its night side at shorter wavelengths.  A hint of this effect is 
provided by a recent upper limit on the day side $K$-band flux (Barnes et al.\ 2007) implying much 
cooler atmospheric temperatures across the deeper near-IR
photosphere than predicted by the no-redistribution model.  Alternately, it could be that the planet possesses an
internal energy source and hence, like Jupiter, simply radiates
more energy than it receives.  One contender for such an energy source is
the tidal dissipation of the orbital eccentricity $e$, which is known to be small
but not zero (Knutson et al.\ 2007a).  However, the predicted dissipation rate (Bodenheimer et al.\ 2003)
based on the Knutson et al.\ (2007a) value is far too small to make this a viable explanation.
Barman (2008) considers the energy budget of this planet in detail.  An additional constraint on the day/night temperature
contrast may come from the recent detection by Swain et al.\ (2008a) of methane (which, at colder temperatures,
becomes the dominant carbon species over CO) in the transmission spectrum of HD~189733b.  While a cool night side
would naturally result in a higher methane abundance, we note that non-equilibrium chemistry as recently explored
by Cooper \& Showman (2006) could also lead to a homogenization of the CO and methane abundances across the planet.

The estimate of the eclipse depth in the Knutson et al.\ (2007a) 8.0 \micron\ data was
$0.3381 \pm 0.0055$\%, which is 2.3$\sigma$ less deep that the value we present here (Table~1). 
This raises the unnerving possibility that the dayside emission from the planet may be varying in
time, as discussed by Rauscher et al.\ (2007a).  This would be both a blessing, in that it
would afford  us the opportunity to study weather on this planet, and a curse in that it
would significantly complicate the interpretation of broadband spectra assembled from measurements
across several secondary eclipses.  Star spots cannot account for this change:  Although the star displays a well-documented periodic
brightness variation due to star spots (Winn et al.\ 2007a), we find upon scaling the variation observed at visible 
wavelengths to our 8.0 \micron\ bandpass that star spots would alter the measured eclipse depth by only 
0.0013\%, a negligible effect. 

The observed arrival times of the secondary eclipses provides an exquisite constraint on $e \cos{\omega}$, where
$e$ is the orbital eccentricity and $\omega$ is the argument of periastron (see e.g.\ Charbonneau 2003).  
After accounting for light travel time across the system, Knutson et al.\ (2007a) found that the
secondary eclipse of HD~189733 occurred $120 \pm 24$~s later than the prediction for a circular orbit,
indicating a residual orbital eccentricity of at least $0.10 \pm 0.02$\%.  Our data are not of sufficient
precision to confirm this offset.  We find that in 5 of the 6 bandpasses, the eclipse center occurs at a time
that is consistent with the expectation of a circular orbit (assuming the value of $P$ from Winn et al.\ 2007a and
the value of $T_c$ from Knutson et al.\ 2007a; see Table~1).  Intriguingly, the eclipse at 3.6~\micron\
was observed to arrive $5.6 \pm 0.8$~minutes later than expected.  Since this offset is chromatic, is
surely isn't due to orbital eccentricity, but instead may arise from strong brightness variations across
the visible hemisphere of the planet (Williams et al.\ 2006; Rauscher et al.\ 2007b).  Despite the seemingly high
significance of this offset, we caution that this band is both the one with the largest instrumental corrections
arising from intra-pixel sensitivity variations (Figure~1) and the largest residual systematic noise
(Figure~2).  Nonetheless, it may be that with greater cadence and more photons, one could detect the
perturbations at ingress and egress that result from a brightness variation across the dayside hemisphere.
By electing to observe in a single band and obviating the need to continually re-point the telescope as
was necessary for the current data, one could realize an improvement of more than a factor of 4 in both cadence
and total photons.  Importantly, such observations could be accomplished during the post-cryogenic phase
of the {\it Spitzer} Mission.

One of the salient features of the population of hot Jupiter exoplanets is its diversity.  These
objects occupy what might first appear to be a rather limited range of orbital phase space.
However, much hard work over the past 8 years has found 28 such planets that transit, providing 
precise constraints on their masses and radii. The scatter in their physical properties is
astounding, and has challenged a community of theorists that must, after all, attempt to unify the bestiary.  
Similarly, with our first detailed comparative studies of the atmospheres of these exoplanets, we are once
again confronted by the differences amongst objects in seemingly similar environments.  This diversity
is both shocking and invigorating to the theoretical enterprise.  We have only one more year to gather
observations longward of 4.5 \micron, at which time {\it Spitzer} will exhaust its cryogen and 
the unique window into the atmospheres of these exoplanets will close.  Fortunately,
we have learned that much of the key information regarding these atmospheres will be, in principle,
contained in the two bandpasses afforded by the post-cryogenic phase of the {\it Spitzer} Mission.
If it is approved, the {\it Spitzer} Warm Mission will therefore be the pre-eminent tool for investigating
the atmospheres of these distant worlds until the launch of {\it JWST} in the middle of the next decade.

\acknowledgments
This work is based on observations made with the {\it Spitzer Space Telescope}, 
which is operated by the Jet Propulsion Laboratory, California Institute of Technology 
under a contract with NASA. Support for this work was provided by NASA through an award 
issued by JPL/Caltech.  HAK was supported by a National Science Foundation Graduate Research 
Fellowship.

Facilities: \facility{Spitzer(IRAC, IRS, MIPS)}.

\newpage

\begin{deluxetable}{lrrrrcrrrrr}
\tabletypesize{\scriptsize}
\tablecaption{Best-Fit Eclipse Depths and Times \label{eclipse_depths}}
\tablewidth{0pt}
\tablehead{
\colhead{$\lambda$ (\micron)} & \colhead{~~Relative Eclipse Depth}  & \colhead{Center of Eclipse (HJD)} & \colhead{O$-$C (min.)\tablenotemark{b}}}
\startdata
3.6 & $0.00256\pm0.00014$ & $2453699.28340\pm0.00044$ & $5.6\pm0.8$\\
4.5 & $0.00214\pm0.00020$ & $2453699.27965\pm0.00087$ & $0.2\pm1.3$\\
5.8 & $0.00310\pm0.00034$ & $2453699.28281\pm0.00153$ & $4.7\pm2.2$\\
8.0 & $0.00391\pm0.00022$ & $2453699.28013\pm0.00102$ & $0.9\pm1.5$\\
16\tablenotemark{a} & $0.00519\pm0.00020$ & $2453692.62302\pm0.00056$ & $-1.1\pm0.9$\\
24 & $0.00598\pm0.00038$ & $2453685.96845\pm0.00094$ & $0.5\pm0.7$\\
\enddata
\tablenotetext{a}{Our analysis of data from \citet{deming2006a}.}
\tablenotetext{b}{$Observed - Calculated$ times of center of eclipse, with the predicted based on the time of center of transit $T_c$ from \citet{knutson07a} and the period $P$ from \citet{winn07a}.}
\end{deluxetable}

\end{document}